\documentclass{PoS}
\usepackage{amssymb,amsbsy,amsmath,amsfonts,bm}
\usepackage[numbers]{natbib}
\usepackage{epstopdf}

\title{Octet baryon masses and sigma terms in covariant baryon chiral perturbation theory}

\ShortTitle{Octet baryon masses and sigma terms}

\author{\speaker{Xiu-Lei~Ren}
\\
       School of Physics and Nuclear Energy Engineering \& International Research Center for Nuclei and
Particles in the Cosmos, Beihang University, Beijing 100191, China \\
       Institut de Physique Nucl\'{e}aire, CNRS-IN2P3, Univ. Paris-Sud,
       Universit\'{e} Paris-Saclay, 91406 Orsay Cedex, France\\
       E-mail: \email{xiulei.ren@buaa.edu.cn}}

\author{Li-Sheng Geng\\
       School of Physics and Nuclear Energy Engineering \& International Research Center for Nuclei and
Particles in the Cosmos, Beihang University, Beijing 100191, China \\
       E-mail: \email{lisheng.geng@buaa.edu.cn}}

\author{Jie Meng\\
      State Key Laboratory of Nuclear Physics and Technology, School of Physics, Peking University, Beijing 100871, China\\
       School of Physics and Nuclear Energy Engineering \& International Research Center for Nuclei and
Particles in the Cosmos, Beihang University, Beijing 100191, China \\
Department of Physics, University of Stellenbosch, Stellenbosch 7602, South Africa\\
       E-mail: \email{mengj@pku.edu.cn}}

\abstract{
We report on a recent study of the ground-state octet baryon masses and sigma terms in covariant baryon chiral perturbation theory with the extended-on-mass-shell scheme up to next-to-next-to-next-to-leading order. To take into account lattice QCD artifacts, the finite-volume corrections and finite lattice spacing discretization effects are carefully examined.
We perform a simultaneous fit of all the $n_f = 2+1$ lattice octet baryon masses and find that the various lattice simulations are consistent with each other. Although the finite lattice spacing discretization effects up to $\mathcal{O}(a^2)$ can be safely ignored, the finite volume corrections cannot even for configurations with $M_\phi L>4$. As an application, we predict the octet baryon sigma terms using the Feynman-Hellmann theorem. In particular, the pion- and strangeness-nucleon sigma terms are found to be $\sigma_{\pi N} = 55(1)(4)$ MeV and $\sigma_{sN} = 27(27)(4)$ MeV, respectively.
}

\FullConference{The 8th International Workshop on Chiral Dynamics, CD2015\\
		29 June 2015 - 03 July 2015\\
		Pisa, Italy}

\begin{document}
\section{Introduction}	
Recently, the lowest-lying octet baryon masses have been simulated by various lattice quantum chromodynamics (LQCD) collaborations~\cite{Durr:2008zz,Alexandrou:2009qu,Aoki:2008sm,Aoki:2009ix,WalkerLoud:2008bp,Lin:2008pr,Bietenholz:2010jr,Bietenholz:2011qq,Beane:2011pc}. Because the limitation of computing resources, most lattice QCD
simulations still have to employ larger than physical light-quark masses~\footnote{Recently, LQCD simulations with physical light-quark masses have
become available (see, e.g., Refs.~\cite{Torrero:2014pxa,Abdel-Rehim:2015owa}).}, finite lattice volumes and lattice spacings.
Therefore, one has to perform the multiple extrapolations of lattice data to the physical point with physical quark masses $(m_q\rightarrow m_q^\mathrm{phys.})$, to the infinite space ($L\rightarrow \infty$), and to the continuum ($a\rightarrow 0$).

Chiral perturbation theory (ChPT)~\cite{Weinberg:1978kz,Gasser:1983yg}, as an effective field theory of low-energy QCD, provides a model independent framework to study the light-quark mass dependence (chiral extrapolation)~\cite{Leinweber:2003dg,Bernard:2003rp,Procura:2003ig,Bernard:2005fy,Young:2009zb, MartinCamalich:2010fp, Semke:2011ez, Bruns:2012eh}, finite-volume corrections (FVCs)~\cite{Gasser:1986vb,Gasser:1987zq,AliKhan:2003ack,Colangelo:2010ba,Geng:2011wq,Lutz:2014oxa}, and continuum extrapolations~\cite{Beane:2003xv,Arndt:2004we,Tiburzi:2005vy} to LQCD data.

In this talk, we present a nice interplay between lattice QCD and ChPT to study the baryon masses.
We calculate the lowest-lying octet baryon masses in covariant baryon chiral perturbation theory (BChPT) with the extended-on-mass-shell (EOMS) scheme~\footnote{See, e.g., Ref.~\cite{Geng:2013xn} for a review of recent developments of EOMS BChPT in the SU(3) sector.} up to next-to-next-to-next-to-leading order (N$^3$LO). Through a systematic study of all the $n_f=2+1$ LQCD data, the light-quark mass dependence on baryon masses is explored, and the finite volume effects and lattice spacing discretization effects are also evaluated self-consistently. By utilizing the Feynman-Hellmann theorem, the octet baryon sigma terms are predicted.

\section{Theoretical Framework}

\begin{figure}[b!]
  \centering
  \includegraphics[width=0.9\textwidth]{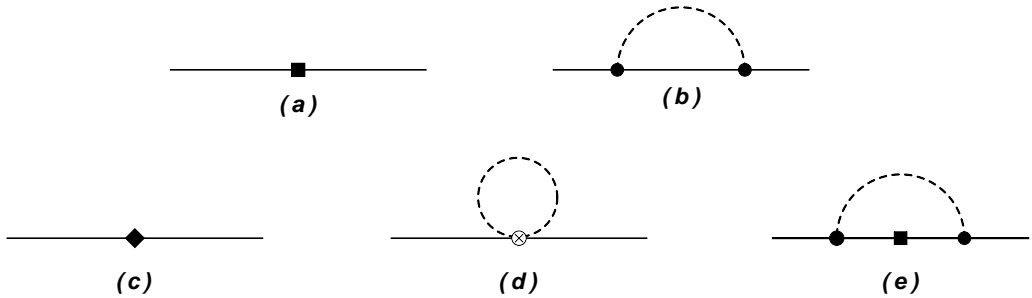}\\
  \caption{Feynman diagrams contributing to the octet baryon masses up to $\mathcal{O}(p^4)$ in EOMS BChPT.}
  \label{Fig:feynman}
\end{figure}

In the continuum space-time, the chiral expansion of the lowest-lying octet baryon masses up to N$^3$LO can be written as
\begin{eqnarray}\label{Eq:N3LO}
  m_B &=& m_0 + m_B^{(2)} + m_B^{(3)} + m_B^{(4)},
\end{eqnarray}
where $m_0$ is the chiral limit octet baryon mass, $m_B^{(2)}$, $m_B^{(3)}$, and $m_B^{(4)}$ correspond to the $\mathcal{O}(p^2)$, $\mathcal{O}(p^3)$, and $\mathcal{O}(p^4)$ contributions from EOMS BChPT, respectively. After calculating Feynman diagrams of Fig.~\ref{Fig:feynman} and removing the power-counting breaking terms, one can obtain the explicit expressions of octet baryon masses:
\begin{eqnarray}\label{Eq:exN3LO}
  m_B &=&  m_0 + \sum\limits_{\phi=\pi, K}\xi_{B,\phi}^{(a)}M_\phi^2 + \sum\limits_{\phi=\pi, K, \eta} \xi_{B,\phi}^{(b)} H_\mathrm{EOMS}^{(b)} \nonumber\\
  && + \sum\limits_{\phi_1,\phi_2}\xi_{B,\phi_1,\phi_2}^{(c)} M_{\phi_1}^2M_{\phi_2}^2 + \sum\limits_{\phi=\pi, K, \eta}\xi_{B,\phi}^{(d)} H_\mathrm{EOMS}^{(d)} + \sum\limits_{\phi=\pi, K, \eta}\xi_{B,\phi}^{(e)} H_\mathrm{EOMS}^{(e)},
\end{eqnarray}
where the coefficients $\xi$ and the loop functions $H_\mathrm{EOMS}$ can be found in Ref.~\cite{Ren:2012aj}.

Because lattice QCD simulations are performed in a finite hypercube, the momenta of virtual particles are discretized. One has to replace a momentum integral by a finite sum of discretized momenta,
\begin{equation}
  \int_{-\infty}^{+\infty} d k ~~~ \rightarrow \sum\limits_{n=-N+1}^{N}\frac{2\pi n}{L},
\end{equation}
with $N=L/(2a)$ (assuming periodical boundary conditions). Thus, lattice results are different from those of infinite space-time. The difference is termed as FVCs: $\delta H_\mathrm{FVCs} = H_B(L)- H_B(\infty)$, where $H_B(L)$ and $H_B(\infty)$ denote the integrals calculated in a finite hypercube and in infinite space-time. With the commonly employed lattice box size $L$ ($3\sim 5$ fm), FVCs cannot be negligible. Therefore, one should obtain the octet baryon masses in a finite-volume box with the replacement of loop functions $H_\mathrm{EOMS}^{(b,d,e)}$ in Eq.~(\ref{Eq:exN3LO}) by
$H_\mathrm{EOMS}^{(b,d,e)}+\delta H_\mathrm{FVCs}^{(b,d,e)}$.

%In Ref.~\cite{Ren:2012aj}\cite{Ren:2013dzt}, we give the corresponding expressions of FVCs related to octet baryon masses.

Furthermore, in order to perform the continuum extrapolation of LQCD simulations to evaluate the discretization effects, one can first write down the Symanzik's effective filed theory~\cite{Symanzik:1983dc,Sheikholeslami:1985ij}:
\begin{equation}
  S_\mathrm{eff} = S_0 + aS_1 + a^2 S_2 + \cdots,
\end{equation}
where $S_0$ is the normal (continuum) QCD action, and $S_{2,3}$ are introduced to include the discretization effects of LQCD. According to Ref.~\cite{Bar:2003mh}, one has the expansion parameters:
\begin{equation}
  \epsilon^2 \sim \frac{m_q}{\Lambda_\chi} \sim a \Lambda_\chi,
\end{equation}
where $\epsilon$ denotes a generic small quantity and $\Lambda_\chi \simeq 1$ GeV denotes the typical chiral symmetry breaking scale. We constructed the SU(3) chiral Lagrangians for Wilson fermion to study the finite lattice spacing effects on the octet baryon masses up to $\mathcal{O}(a^2)$, which can be written as
\begin{equation}\label{Eq:Aeffects}
  m_B^{(a)} = m_B^{\mathcal{O}(a)} + m_B^{\mathcal{O}(am_q)} + m_B^{\mathcal{O}(a^2)},
\end{equation}
where the expressions of $m_B^{\mathcal{O}(a), \mathcal{O}(am_q), \mathcal{O}(a^2)}$ can be found in Ref.~\cite{Ren:2013wxa}.

Here we want to mention that there are $19$ unknown low-energy constants (LECs), $m_0$, $b_0$, $b_D$, $b_F$, $b_{1, \cdots,8}$, $d_{1,\cdots,5,7,8}$, needed to be fixed in EOMS BChPT at $\mathcal{O}(p^4)$. Furthermore, including the finite lattice spacing effects [Eq.~(\ref{Eq:Aeffects})], one has to introduce $4$ more  combinations of the unknown LECs in the study of LQCD data based on the $\mathcal{O}(a)$-improved Wilson fermion~\cite{Ren:2013wxa}. The details of the studies can be found in Refs.~\cite{Ren:2012aj,Ren:2013wxa}.

\section{Systematic study of lattice QCD data}

In order to determine all the LECs and test the consistency of the current LQCD simulations, we perform a simultaneous fit to all the publicly available $n_f=2+1$ LQCD data from the PACS-CS~\cite{Aoki:2008sm}, LHPC~\cite{WalkerLoud:2008bp}, QCDSF-UKQCD~\cite{Bietenholz:2011qq},  HSC~\cite{Lin:2008pr}, and NPLQCD~\cite{Beane:2011pc} collaborations. To ensure that the N$^3$LO BChPT stay in its applicability range, fitted LQCD data are limited to those satisfying $M^2_{\pi}< 0.25$ GeV$^2$ and $M_{\phi}L>4$.

\begin{table}[t]
\centering
\caption{Values of the LECs and the corresponding $\chi^2/\rm{d.o.f.}$ from the best fits.
 We have performed fits  to the LQCD and experimental data at $\mathcal{O}(p^2)$, $\mathcal{O}(p^3)$, and $\mathcal{O}(p^4)$, respectively.}
\label{Tab:fitcoef}
~~\\[0.1em]
\begin{tabular*}{0.8\textwidth}{@{\extracolsep{\fill}}cccc}
\hline\hline
         & NLO & NNLO  &  N$^3$LO \\
\hline
  $m_0$~[MeV]         & $900(6)$      &   $767(6)$     &  $880(22)$        \\
  $b_0$~[GeV$^{-1}$]  &$-0.273(6)$    &  $-0.886(5)$   &  $-0.609(19)$    \\
  $b_D$~[GeV$^{-1}$]  &$0.0506(17)$   &  $0.0482(17)$  &  $0.225(34)$      \\
  $b_F$~[GeV$^{-1}$]  &$-0.179(1)$    &  $-0.514(1)$   &  $-0.404(27)$     \\
  $b_1$~[GeV$^{-1}$]  & --            &  --            &  $0.550(44)$      \\
  $b_2$~[GeV$^{-1}$]  & --            &  --            &  $-0.706(99)$       \\
  $b_3$~[GeV$^{-1}$]  & --            &  --            &  $-0.674(115)$      \\
  $b_4$~[GeV$^{-1}$]  & --            &  --            &  $-0.843(81)$       \\
  $b_5$~[GeV$^{-2}$]  & --            &  --            &  $-0.555(144)$    \\
  $b_6$~[GeV$^{-2}$]  & --            &  --            &  $0.160(95)$      \\
  $b_7$~[GeV$^{-2}$]  & --            &  --            &  $1.98(18)$      \\
  $b_8$~[GeV$^{-2}$]  & --            &  --            &  $0.473(65)$     \\
  $d_1$~[GeV$^{-3}$]  & --            &  --            &  $0.0340(143)$   \\
  $d_2$~[GeV$^{-3}$]  & --            &  --            &  $0.296(53)$      \\
  $d_3$~[GeV$^{-3}$]  & --            &  --            &  $0.0431(304)$   \\
  $d_4$~[GeV$^{-3}$]  & --            &  --            &  $0.234(67)$     \\
  $d_5$~[GeV$^{-3}$]  & --            &  --            &  $-0.328(60)$    \\
  $d_7$~[GeV$^{-3}$]  & --            &  --            &  $-0.0358(269)$  \\
  $d_8$~[GeV$^{-3}$]  & --            &  --            &  $-0.107(32)$     \\
  \hline
$\chi^2$/d.o.f. & $11.8$ & $8.6$  & $1.0$ \\
\hline\hline
\end{tabular*}
\end{table}

In Table~\ref{Tab:fitcoef}, we tabulate the values of LECs from the fit of N$^3$LO BChPT. For comparison, we have fitted lattice data using the $\mathcal{O}(p^2)$ and $\mathcal{O}(p^3)$ mass formulas. The corresponding values of LECs $b_0$, $b_D$, $b_F$, and $m_0$ are also given in Table~\ref{Tab:fitcoef}. With the decrease of fit-$\chi^2/\mathrm{d.o.f.}$, we found that the EOMS BChPT shows a good description of LQCD and experimental data with order-by-order improvement. Up to N$^3$LO, the $\chi^2/{\rm d.o.f.}$ is about $1.0$, which indicates that the lattice simulations from these five collaborations are consistent with each other~\footnote{This does not seem to be the case for LQCD simulations of the ground-state decuplet baryon masses~\cite{Ren:2013oaa}.}, although their setups are very different. Finally, it is essential to point out that including FVCs is important to understand LQCD results in ChPT at N$^3$LO. Without FVCs taken into account, the best fit to lattice data yields $\chi^2/\mathrm{d.o.f.}\sim 1.9$. Furthermore, in Ref.~\cite{Ren:2013dzt}, we also performed a systematic study of virtual decuplet contributions to the octet baryon masses, and found that their effects on the chiral extrapolation and finite-volume corrections are very small.

\begin{figure}[t]
  \centering
  \includegraphics[width=\textwidth]{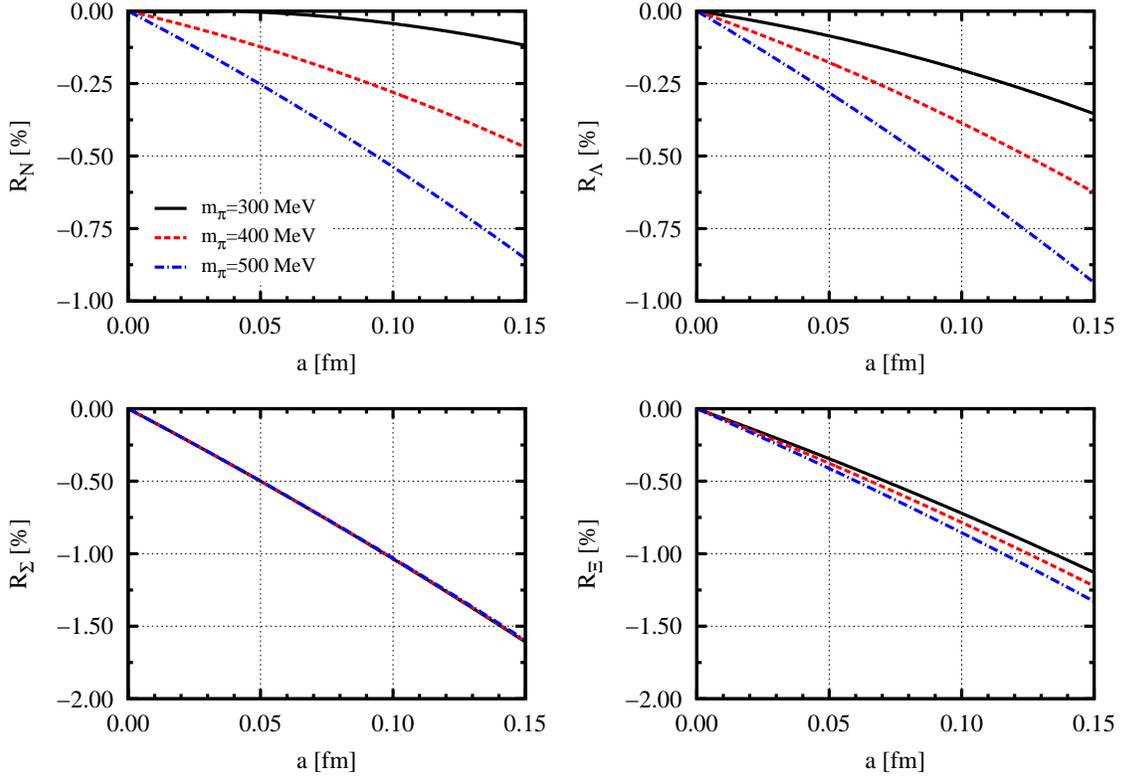}\\
  \caption{Finite lattice spacing effects on octet baryon masses, $R_B=m_B^{(a)}/m_B$, as functions of lattice spacing $a$ for $M_\pi=0.3$, $0.4$, and $0.5$ GeV, respectively.}
  \label{Fig:ampi-la}
\end{figure}

In Ref.~\cite{Ren:2013wxa}, we studied the discretization effects on the ground-state octet baryon masses by analyzing the latest $n_f=2+1$ $\mathcal{O}(a)$-improved LQCD data of the PACS-CS, QCDSF-UKQCD, HSC and NPLQCD collaborations. In Fig.~\ref{Fig:ampi-la}, we show the evolution of discretization effects as a function of the lattice spacing for three different pion masses. It is seen that the discretization effects increase almost linearly with increasing lattice spacing for fixed pion mass. For the fixed lattice spacing, they increase with increasing pion mass as well. We also found that the finite lattice spacing effects are at the order of $1-2$\% of octet baryon masses for lattice spacings up to $0.15$ fm and the pion mass up to $500$ MeV, which is in agreement with other LQCD studies.

\section{Octet baryon sigma terms}

Nucleon sigma terms, $\sigma_{\pi N}$ and $\sigma_{s N}$, as emphasized by H. Leutwyler~\cite{Leutwyler:2015jga}, play an important role in understanding the composition of nucleon mass, and are also important in the study of dark matter searches. In Ref.~\cite{Ren:2014vea}, we use the
Feynman-Hellmann theorem to calculate the octet baryon sigma terms,
\begin{eqnarray}
  \sigma_{\pi B} &=& m_l\langle B|\bar{u}u + \bar{d}d|B\rangle \equiv m_l \frac{\partial m_B}{\partial m_l},\\
  \sigma_{\pi B} &=& m_s\langle B|\bar{s}s|B\rangle \equiv m_s \frac{\partial m_B}{\partial m_s},
\end{eqnarray}
where $m_l\equiv m_u=m_d$ is the up or down quark mass with exact isospin symmetry, $m_s$ is the strange quark mass, and $m_B$ is the chiral expansions of octet baryon masses up to N$^3$LO [Eq.~(\ref{Eq:exN3LO})].

In order to obtain an accurate determination of sigma terms, a careful examination of the LQCD data is essential, since not all of them are of the same quality though they are largely consistent with each other. In Ref.~\cite{Ren:2014vea}, we only selected high statistic lattice data from the PACS-CS, LHPC and QCDSF-UKQCD collaborations. We also took into account the scale setting effects of LQCD simulations (mass dependent scale (MDS) setting and mass-independent scale (MIS) setting) and studied systematic uncertainties from truncating chiral expansions. Furthermore, strong-interaction isospin breaking effects to the baryon masses were for the first time employed to better constrain the relevant LECs up to N$^3$LO.

\begin{table}[t!]
  \centering
  \caption{Predicted pion- and strangeness-sigma terms of octet baryons (in units of MeV) with N$^3$LO BChPT.}
  \label{Tab:N3LOsigma}
~~\\[0.1em]
\begin{tabular*}{0.8\textwidth}{@{\extracolsep{\fill}}ccccc}
  \hline\hline
    & \multicolumn{2}{c}{MIS} & MDS \\
  \cline{2-3}
      & $a$ fixed & $a$ free &  \\
  \hline
  $\sigma_{\pi N}$     & $55(1)(4)$ & $54(1)$ & $51(2)$ \\
  $\sigma_{\pi \Lambda}$  & $32(1)(2)$ & $32(1)$ & $30(2)$  \\
  $\sigma_{\pi \Sigma}$    & $34(1)(3)$ & $33(1)$ & $37(2)$  \\
  $\sigma_{\pi \Xi}$      & $16(1)(2)$ & $18(2)$ & $15(3)$  \\
  \hline
  $\sigma_{s N}$        & $27(27)(4)$  & $23(19)$  & $26(21)$  \\
  $\sigma_{s \Lambda}$  & $185(24)(17)$ & $192(15)$ & $168(14)$ \\
  $\sigma_{s \Sigma}$  & $210(26)(42)$ & $216(16)$ & $252(15)$ \\
  $\sigma_{s \Xi}$     & $333(25)(13)$ & $346(15)$ & $340(13)$ \\
  \hline\hline
\end{tabular*}
\end{table}

\begin{figure}[t!]
  \centering
  \includegraphics[width=0.66\textwidth]{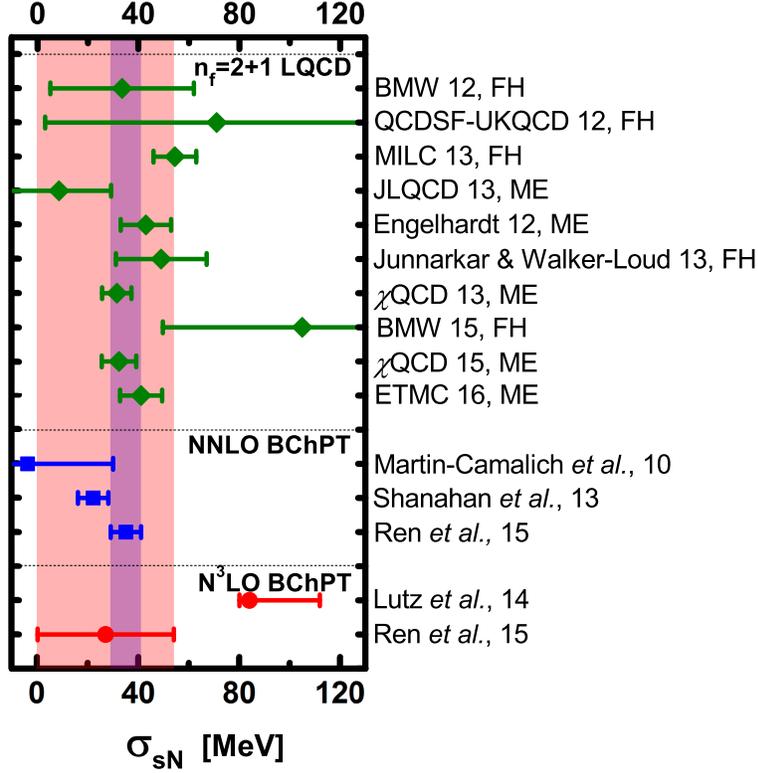}\\
  \caption{Strangeness-nucleon sigma term determined from different studies.
The purple and pink bands are our NNLO and N$^3$LO results, respectively.
   Data points are taken from the following references: BMW~\cite{Durr:2011mp,Durr:2015dna}, QCDSF-UKQCD~\cite{Horsley:2011wr}, MILC~\cite{Freeman:2012ry}, and Junnarkar \& Walker-Loud~\cite{Junnarkar:2013ac} using the Feynman-Hellmann (FH) theorem; JLQCD~\cite{Oksuzian:2012rzb}, Engelhardt~\cite{Engelhardt:2012gd}, $\chi$QCD~\cite{Gong:2013vja,Yang:2015uis}, and ETMC~\cite{Abdel-Rehim:2016won} calculating the scalar matrix elements (ME); Martin-Camalich {\it et al.}~\cite{MartinCamalich:2010fp}, Shanahan {\it et al.}~\cite{Shanahan:2012wh}, Lutz {\it et al.}~\cite{Lutz:2014oxa}, Ren {\it et al.}~\cite{Ren:2014vea}.
  }
  \label{Fig:sigma_s}
\end{figure}

In Table~\ref{Tab:N3LOsigma}, we list the predicted baryon sigma terms. Our results are consistent with each other within uncertainties, and the scale setting effects on the sigma terms seem to be small. Therefore, we take the central values from the fit to the mass independence $a$ fixed LQCD simulations as our final results.
Our obtained nucleon sigma terms are $\sigma_{\pi N}=55(1)(4)$ MeV and $\sigma_{sN}=27(27)(4)$ MeV, which are consistent with recent LQCD and BChPT studies. Furthermore, our predicted strangeness-nucleon sigma term is shown in Fig.~\ref{Fig:sigma_s}. It is clear that the NNLO result has a much smaller uncertainty compared to the N$^3$LO. Therefore, further high statistic LQCD baryon masses, especially for lattice simulations with different strange quark masses, are necessary to reduce the uncertainty of $\sigma_{s N}$.%~\footnote{After our work~\cite{Ren:2014vea} was published, the BMW~\cite{{Durr:2015dna}}, $\chi$QCD~\cite{Yang:2015uis} and ETMC~\cite{Abdel-Rehim:2016won} collaborations reported update determinations of the sigma terms at the physical point.}

\section{Conclusions}

We have studied the lowest-lying octet baryon masses in EOMS BChPT up to N$^3$LO. The unknown low-energy constants are determined by a simultaneous fit to the latest $n_f=2+1$ LQCD simulations, and it is shown that the LQCD results are consistent with each other, though their setups are quite different.
The finite-volume corrections and finite-lattice spacing discretization effects on the LQCD baryon masses have been evaluated as well. We find that their effects
are of similar size  but finite volume corrections are more important to better constrain the LECs and to reduce the $\chi^2/\mathrm{d.o.f.}$.

Using the Feynman-Hellmann theorem, we have performed an accurate determination of the nucleon sigma terms, focusing on the uncertainties from the lattice scale setting method and chiral expansions. Our predictions are $\sigma_{\pi N}=55(1)(4)$ MeV and $\sigma_{sN}=27(27)(4)$ MeV, which are
 consistent with most of the recent LQCD and BChPT studies. However, further LQCD simulations
 are needed to reduce the uncertainty of the nucleon strangeness-sigma term.

\section*{Acknowledgments}
X.-L.R acknowledges financial support from the China Scholarship Council.
This work was partly supported by the National
Natural Science Foundation of China under Grants No. 11375024, No. 11522539, and No. 11411130147.

%\begin{thebibliography}{000} %for 3 digits
%\begin{thebibliography}{00}  %for 2 digits

\end{document}